\documentclass[twocolumn,aps,showpacs,prl,floatfix]{revtex4}%
\usepackage{epsfig}
\usepackage{dcolumn}
\usepackage{bm}
\usepackage{amsmath}
\usepackage{amsfonts}
\usepackage{amssymb}
\usepackage{graphicx}
\newcommand{\beq}{\begin{eqnarray}}
\newcommand{\eeq}{\end{eqnarray}}

\begin{document}
\title{Quantum Monte Carlo method for models of molecular nanodevices}
\author{Liliana Arrachea$^{1,2}$
and Marcelo J. Rozenberg$^{2,3}$}
\affiliation{ 
$^{1}$ Max Planck Institut f\"ur Physik komplexer Systeme, Dresden,
N\"othnitzer Str. 38 D-1187, Germany. \\
$^{2}$ Laboratoire de Physique des Solides, CNRS-UMR8502, Universite de Paris-Sud,
Orsay 91405, France.\\
$^{3}$ Departamento de F\'{i}sica, FCEN, Universidad de Buenos Aires, Ciudad Universitaria Pab.1,
1428 Buenos Aires, Argentina.
}

\begin{abstract}
We introduce a quantum Monte Carlo technique to calculate exactly at
finite temperatures
the Green function of a fermionic quantum impurity coupled to
a bosonic field. 
While the algorithm is general, we focus on the single impurity Anderson
model coupled to a Holstein phonon as a schematic model for a molecular
transistor.
We compute the density of states at the impurity
 in a large range of parameters, 
to demonstrate the accuracy and efficiency of the method.
We also obtain the conductance of the impurity model and analyze 
different regimes. The results show that even
 in the case when the effective attractive
phonon interaction is larger than the Coulomb repulsion, a Kondo-like
conductance behavior might be observed.

\end{abstract}

\maketitle

The physics of quantum impurities is attracting a great deal of
attention.
For decades, their study has been a classic subject in the area of strongly correlated
electron systems \cite{hewsonbook}. 
However, with the dramatic recent advances in nanoscale physics
their interest has become paramount to a much broader audience.
Among the most interesting recent achievements in nanometre-scale devices
we find the ability to attach individual molecules to metallic electrodes \cite{vibmol}.
In these systems, the coupling between the discrete internal
degrees of freedom of the molecule and the
macroscopic leads gives rise to new transport channels enabled by resonances
originated in many-body interactions. An important example is the observation
of the physics of the Kondo effect
\cite{konmol}, which were previously seen in 
quantum dots \cite{kondot}.
In general, one expects that the transport through molecular devices
will be affected by the coupling between 
their internal vibration modes and the electrons \cite{vibmol}. 
In addition, one should also consider
the possible electron correlation effects due to Coulomb repulsion
within the molecule
\cite{konmol}.

The solution of quantum impurity models has remained a difficult
task \cite{hewsonbook}, and exact solutions are only possible in very specific
instances with the use of techniques such as the Bethe ansatz.
For general models, however, theoreticians have to relay heavily
on numerical methods. Among those we count the celebrated
Wilson's numerical renormalization group (NRG) and its various
generalizations \cite{hewsonbook}. This method is extremely accurate for
the determination of the low energy physics of the model, but
has limitations for the investigation of intermediate temperatures 
and impurity models with many orbitals.
On the other hand, there is a very efficient
Quantum Monte Carlo (QMC) method introduced
by Hirsch and Fye \cite{hirfy} which 
does not present those
limitations and neither the so called
``negative sign'' problem.

Among models of quantum impurities a most important one is
the single impurity Anderson model (SIAM) Hamiltonian \cite{an}, that has
been crucial for a wide number of physical problems. Essentially it
consists of an atomic orbital (the impurity) with an on-site
repulsive Coulomb interaction that is
hybridized with a band of conduction electrons.

The goal of the present work is twofold: we shall first introduce
a new algorithm which generalizes that of Hirsch and Fye to the 
quantum impurity problem of a SIAM coupled to a general bosonic 
field. In order to test our method we shall consider the particular
case of the SIAM coupled to a Holstein phonon which has been
recently studied with NRG at $T=0$. We shall present the detailed
behavior of the density of states (DOS) of the impurity for various
phonon strengths, local Coulomb interaction and temperatures.
The second goal of our work will be to discuss new results for 
the conductance of this quantum impurity as a function of the
gate voltage, that can be understood under the light of our DOS calculations. 
We shall show the two generic conductance behaviors that occur depending
on the values of the parameters. Our results should be useful
for interpretation of experiments of transport through 
molecular transistors.

The SIAM coupled to a general bosonic
mode is described by the action:
\begin{eqnarray}
{\cal S}& =&\sum_{\sigma} \int d\tau d\tau' 
\overline{c}_{\sigma}(\tau) {G}^{-1}_0(\tau -\tau')
c_{\sigma}(\tau')
\nonumber \\
& & + \int d\tau d\tau' 
\phi(\tau) {\cal \it  D}^{-1}_0(\tau -\tau')
\phi(\tau') +  
\int d \tau U n_{\uparrow} (\tau) n_{\downarrow}(\tau)
\nonumber \\
& & + \lambda\sum_{\sigma}
 \int d\tau \phi(\tau) 
(\overline{c}_{\sigma}(\tau)c_{\sigma}(\tau) -\frac{1}{2}),
\label{action}
\end{eqnarray}
where $\overline{c}_{\sigma}, c_{\sigma}$ are Grassmann variables
for the creation and destruction of an electron with spin $\sigma$
at the impurity and
$\phi$ is the bosonic field. 
The coupling $\lambda$ is the strength of the interaction
between the electrons at the impurity
and the bosonic degree of freedom, 
and $U$ is the Coulomb cost for the double
occupation of the impurity site.
The non-interacting propagators are $D_0(\omega_m)$ for the bosons
and $G_0(\omega_n) = [i\omega_n -V_g+ \Delta(\omega_n)]^{-1}$
for the impurity electrons, being $\omega_m$ and $\omega_n$  bosonic
and fermionic Matsubara frequencies respectively.
The hybridization function
$\Delta(\omega_n)$ describes the coupling to the (given)
conduction band and $V_g$ denotes the gate voltage.

We shall now describe the QMC
algorithm.
As usual one
performs a Trotter breakup of imaginary time into
$M$ time slices $\tau_l$ of length $\Delta \tau = \beta / M$
to discretize the action (\ref{action})
and introduce a discrete Hubbard-Stratonovich transformation
with auxiliary Ising-like fields $s_l, l=0, \ldots , M-1$.

This renders the action quadratic in the fermionic variables
which can be integrated out.
The resulting partition function is
\begin{eqnarray}
Z&=& \mbox{Tr}_{\{s_l\}} \int {\cal D} \phi
{\mbox Det}[G^{-1}_{\uparrow}(\tau_l,\tau_{l'})] 
{\mbox Det}[G^{-1}_{\downarrow}(\tau_l,\tau_{l'})] \nonumber \\
& &\exp{\{\frac{1}{\beta}\sum_{m} \overline{\phi}^*_m 
{\cal \it D}^{-1}_0(i\omega_m) \overline{\phi}_m 
+ \lambda \overline{\phi}_0\}},
\label{part}
\end{eqnarray}
where $\overline{\phi}_m$ are the Fourier transform components of the
bosonic field that obey $\overline{\phi}_m=\overline{\phi}^{*}_{-m}$.
The $s_l$ and $\phi_m$ dependent electron Green functions are
\cite{noteqmc}
\begin{equation}
G^{-1}_{\sigma s_l \phi}(\tau_l,\tau_{l'}) ={G}^{-1}_0 (\tau_l-\tau_{l'})
+\delta_{l,l'}
[ \sigma \lambda_U s_l + \Delta \tau \lambda \phi(\tau_l) ],
\label{green}
\end{equation}
with $\mbox{cosh}\lambda_U= \exp{(\Delta \tau U/2)}$.
A crucial new aspect of our algorithm is the 
choice of a simultaneous
representation in terms of the Ising-like and {\em continuous}
bosonic fields.
Other algorithms
chose to integrate out the bosonic variables to obtain a
time dependent electron density-density interaction which is
then decoupled via a {\em single} continuous Hubbard-Stratonovich
field. That procedure gives up the very nice properties
of the discrete Ising-like fields and should be expected
to have serious autocorrelation problems.
In the present case, each degree of freedom is simulated in
the most effective way. This is evident for the Ising-fields, but
should also be clear for the bosonic fields as our frequency
modes representation is reminiscent of the Fourier acceleration
method. 
`In fact, while the update of the Ising fields is known to lead to
very short autocorrelation times the same is not true for the
bosonic fields. The physical reason is that the most relevant
bosonic configurations are those in which $\phi(\tau)$ have a
slow variation with $\tau$. To generate those configurations
one needs to propose moves that coherently and unbiased
involve all the
M variables $\phi(\tau_l)$. Our algorithm provides a systematic
way to do this by updating the frequency components $\overline{\phi}_m$ of the
bosonic fields. It is important to mention that
the autocorrelation times of  $\overline{\phi}_m$
remain pretty long (typically about a hundred sweeps) but, unlike
what would be updating in time domain,
practically tractable.'

The integrand of (\ref{part}) thus defines the Boltzmann weight 
${\cal W}[s_l;\overline{\phi}_m]$
for the stochastic
evaluation of the trace on the Ising and bosonic fields.
We use a heat-bath algorithm according to which
new configurations $\{s_l;\overline{\phi}_m \}$
are generated and acepted with a probability 
$P({\cal W} \rightarrow  {\cal W}^{\prime})
=  {\cal W}^{\prime}/({\cal W}^{\prime}+{\cal W})$.
The ensuing Markov chain is: (i)
each Ising variable is visited and a 
flip $s_l \rightarrow -s_l$ is proposed.
The acceptance of the move is evaluated as in the usual Hirsch and Fye algorithm 
\cite{hirfy} and the new Green function can be computed in $O(M^2)$ operations.
(ii) every independent complex variable  
$\overline{\phi}_m=(\overline{\phi}^{R}_m,\overline{\phi}^{I}_m)$
is visited and a move 
$\overline{\phi}^{R,I}_m \rightarrow  \overline{\phi}^{R,I}_m +
\delta_m^{R,I}$ is proposed,
with $\delta_m^{R,I} \in [-\Lambda, \Lambda]$ a uniformly
distributed random number.
After Fourier transforming back $\overline\phi$, the update requires
a full matrix inversion (\ref{green}), that is 
$O(M^3)$ operations \cite{note}.
One can verify that, similarly as the original Hirsch and Fye
algorithm, this procedure fulfills detailed balance:
${\cal W} P({\cal W} \rightarrow  {\cal W}^{\prime})=
{\cal W}^{\prime} P({\cal W}^{\prime} \rightarrow  {\cal W})$.

Holstein phonons are bosonic fields defined by its non-interacting correlation
function
${D}_0(i \omega_m)= - 2 \Omega_0 /(\omega_m^2 + \Omega_0^2)$, where $\Omega_0$
is the phonon frequency.
For the electron conduction band that is hybridized with the
impurity we shall take a semicircular density of states, thus
$\Delta(\omega)/2=t_c^2 \rho_0(\omega)=4t_c^2 \sqrt{W^2-\omega^2}/W^2$
with $\rho_0(\omega)$ the conduction electron DOS, $W$ its 
half-bandwidth and $t_c$ is the hybridization coupling.
The choice of a semicircular DOS is made because it would also be appropriate
for our investigation of conductance through the impurity. In fact, 
the same 
$\Delta(\omega)$ also results from
attaching two leads to the impurity in the form of two semi-infinite chains of 
atoms with intersite hopping amplitude $W/2$. This set-up is usually used
for modelling  nanodevices, such as a molecule connected to two metallic
leads.

\begin{figure}
\includegraphics[width=8cm,clip]{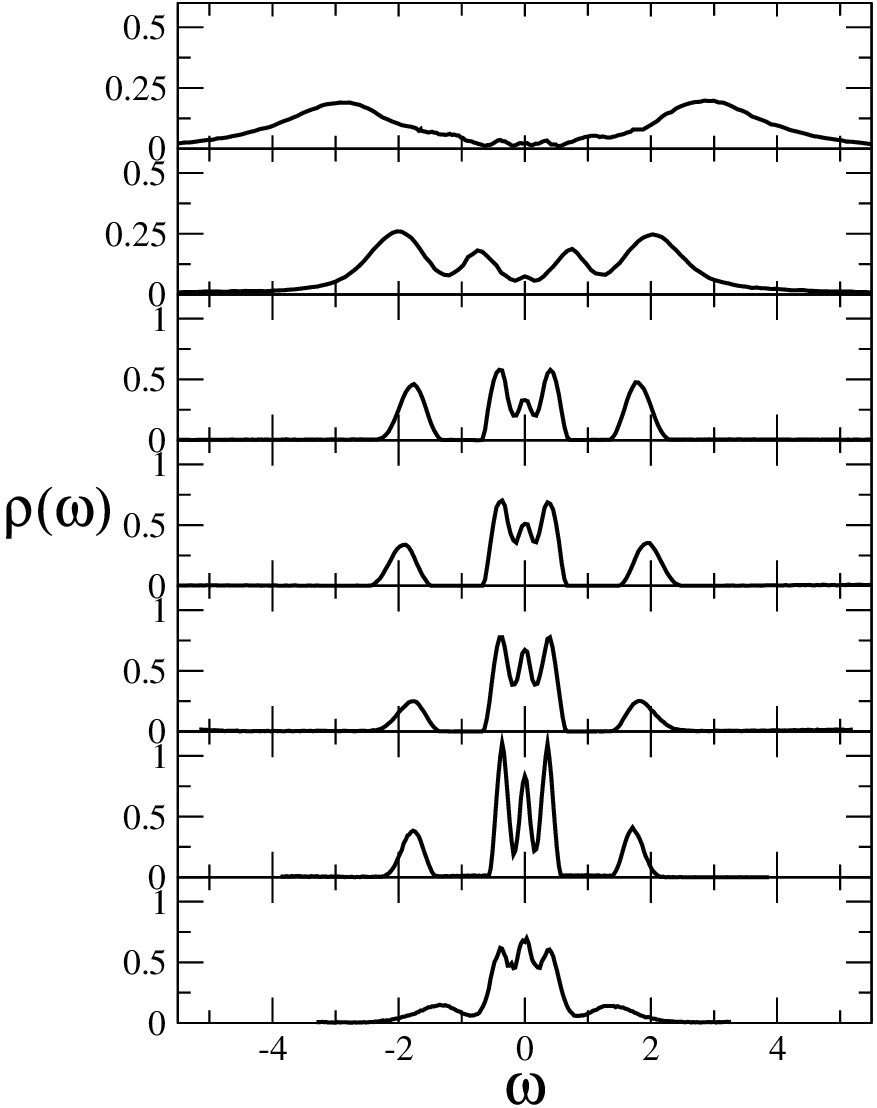}
\caption{DOS for $U=0$, $W=8$, $t_c=1.$ and $T=0.0357$.
Other parameters are: $\lambda=1$, $\Omega_0=0.333,0.5,0.75,1.,
1.5,2.$, and  $\lambda=0.5$, $\Omega_0=2$ (upper to lower panel).
}
\label{fig1}
\end{figure}

The DOS at the impurity $\rho(\omega)=-2\mbox{Im}[G_{\sigma}(\omega)]$
is obtained by 
a Maximum Entropy method for the analytic continuation
of the QMC calculated Green function 
to the real frequency axis. We typically use upto 64 time slices
and 200,000 complete sweeps of the statistical fields. 
We consider first the simpler case of $U=0$ and show in
Fig.\ref{fig1} systematic results for fixed $\lambda$ and increasing
phonon frequency $\Omega_0$. 
For the analysis of the results it is useful to recall 
that within
the static approximation (i.e., taking $D_0(i \omega_m) \approx
D_0(i \omega_0)\delta_{0,m}$), which  becomes more accurate in the
 small $\Omega_0$ limit,
the integration of the bosons leads
to an effective interaction between the electrons
$-U_{eff}=-2\lambda^2/\Omega_0$.
 Since it is attractive, and our model is particle-hole
symmetric, it will favour an empty or doubly occupied impurity
state. Adding or removing and electron from the impurity
has an energy cost of $\sim U_{eff}/2$, which is in fact the
position of the two features in the DOS of the top 
panel of Fig.\ref{fig1} for the smaller $\Omega_0=0.333$.
As $\Omega_0$ is increased and $U_{eff}$ decreases, the DOS
at low frequency has a dramatic enhancement with a strong
quasiparticle peak emerging at $\omega = 0$. This resonance
corresponds to slow charge fluctuations at the impurity
and is the charge counterpart of the Kondo effect.
At higher frequencies a reacher peak structure also evolves,
where the two side peaks of the Kondo-like resonance correspond
to the (dynamically renomalized) $U_{eff}/2$ features, while
the two additional peaks at higher frequency 
$\omega \approx \pm (U_{eff}/2 + \Omega_0)$ correspond to
adding (or removing) a
particle plus a phonon.
When $\Omega_0$ is further increased (diminishing $U_{eff}$), 
charge fluctuations are no longer prevented and the DOS moves
towards that of a simple resonant level model with just
a remanent substructure at frequencies $\omega \sim \Omega_0$.
This basic systematic behavior is consistent with the 
$T=0$ numerical studies using NRG \cite{hew}.

\begin{figure}
\includegraphics[width=8cm,clip]{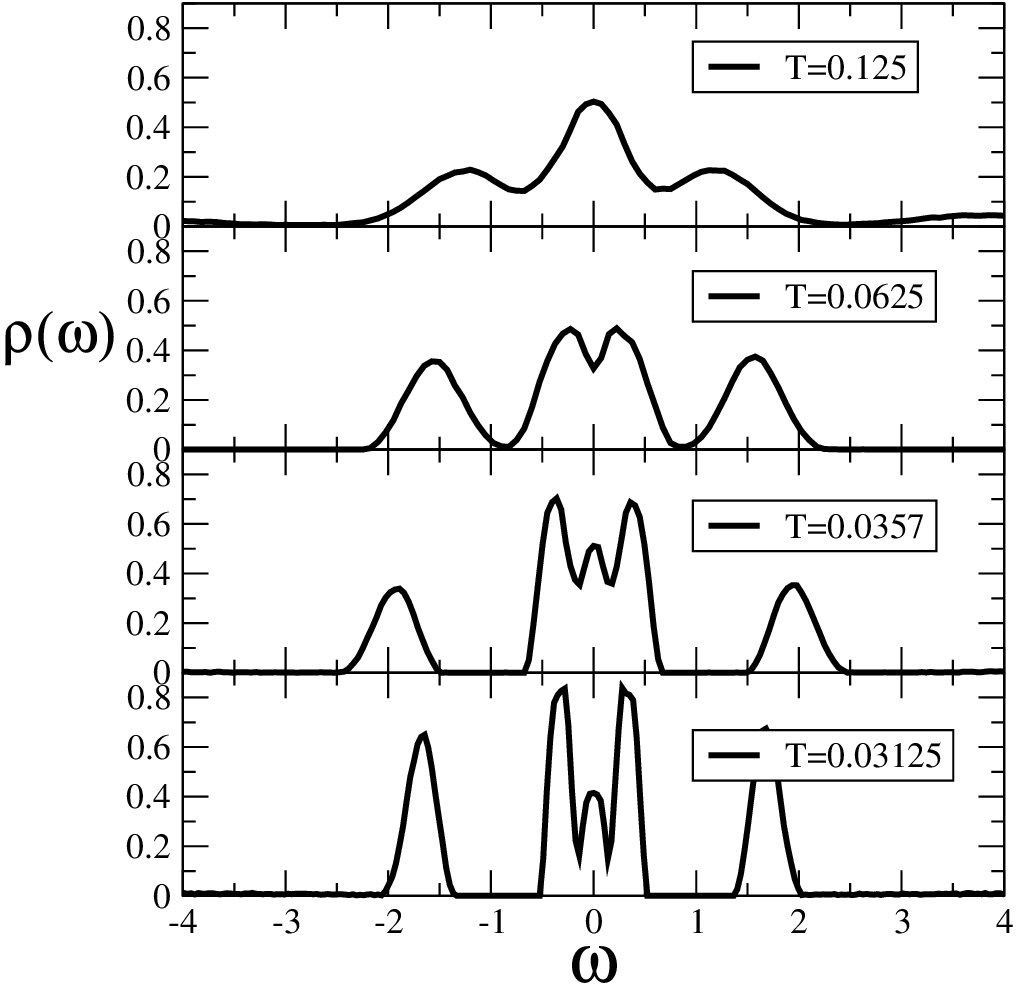}
\caption{Behavior of the DOS with $T$ 
for $\lambda=1$ and $\Omega_0=1$. Other parameters are 
as in Fig.\ref{fig1}.}
\label{fig2}
\end{figure}

We now turn to the evolution of the DOS as a function of $T$ 
shown in
Fig.\ref{fig2}. At high temperatures (upper panel) strong charge fluctuations
dominate and the DOS is that of a resonant model with a strong
and broad feature at $\omega = 0$. In addition, two side peaks are present
at $\omega \sim \pm \Omega_0$ due to excitations to states with an additional
phonon.
Lowering $T$ we observe that the central feature splits-up into two peaks.
This splitting occurs due to the thermal selection of the
more favorable empty and doubly occupied states  
as thermal charge fluctuations decrease.
Finally (two lower panels), 
bellow $T_{\lambda} \sim 0.0625$ (analogous to the
Kondo temperature),  
a strong resonance re-emerges at $\omega=0$ due
to the Kondo-like mechanism in the charge sector.
>From the Kondo model analogy, $T_{\lambda} \propto exp(-U_{eff})$ 
therefore is rapidly suppressed with increasing $\lambda$ or decreasing
$\Omega_0$.
   
\begin{figure}
\includegraphics[width=7cm,clip]{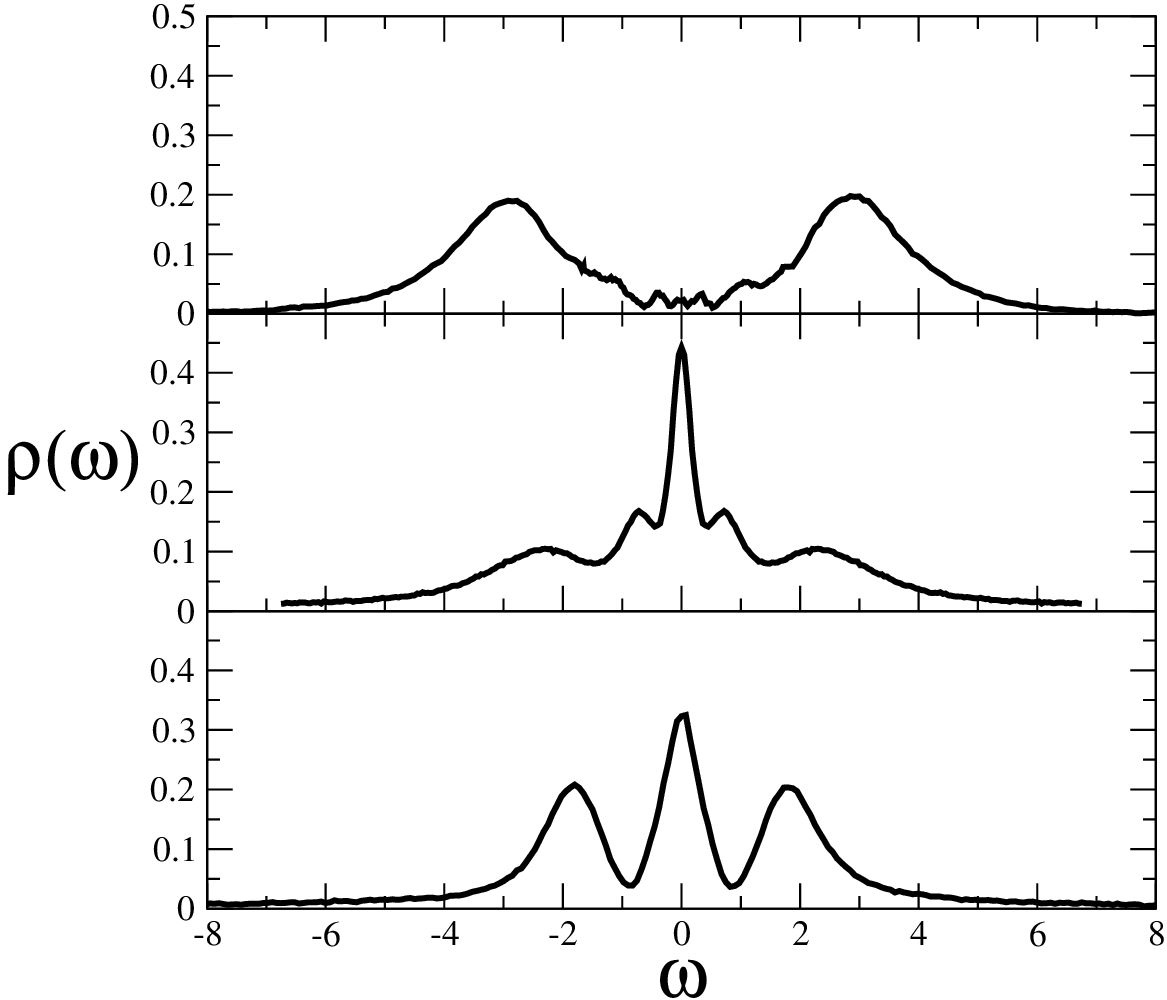}
\caption{Top: DOS without Coulomb repulsion ($U=0$, $\lambda=1$). 
Bottom: DOS without phonons ($U=4$, $\lambda=0$).
Middle: DOS with Coulomb repulsion and phonons ($U=4$, $\lambda=1$).
The phonon frequency is $\Omega_0=0.333$, $T=0.0357$ (top) and $T=0.0625$
(middle and bottom).
}
\label{fig3}
\end{figure}
We now switch on the interaction $U$ and observe
the interplay between the usual Kondo effect and the vibrational modes
(Fig.\ref{fig3}). Without phonons, ie, $\lambda =0$
(bottom) the DOS consists of a central Kondo
resonance and side features at $\pm U/2$. The familiar Kondo peak is due to slow
fluctuations between the spin-up and spin-down states within
the charge sector of single occupation. Upon coupling to the 
vibrational modes the effective electronic interaction
of the impurity becomes dynamical and strongly renormalized
(screened). Within the static approximation, its value
becomes $U_{dyn}= U - U_{eff} = U - 2 \lambda^2/\Omega_0$.
The effective phonon interaction can even render the interaction
attractive as in the case of superconductivity.
The effect on the impurity DOS is quite notable (middle), renormalizing
the side peaks to $\omega \sim \pm U_{dyn}/2$ and adding higher frequency
peaks due to additional phonon excitation.
A central Kondo-like resonance remains, however, it is now due to
a rather complicated many body state involving both slow spin and charge
fluctuations.

\begin{figure}[ht]
\includegraphics[width=7cm,clip,angle=0]{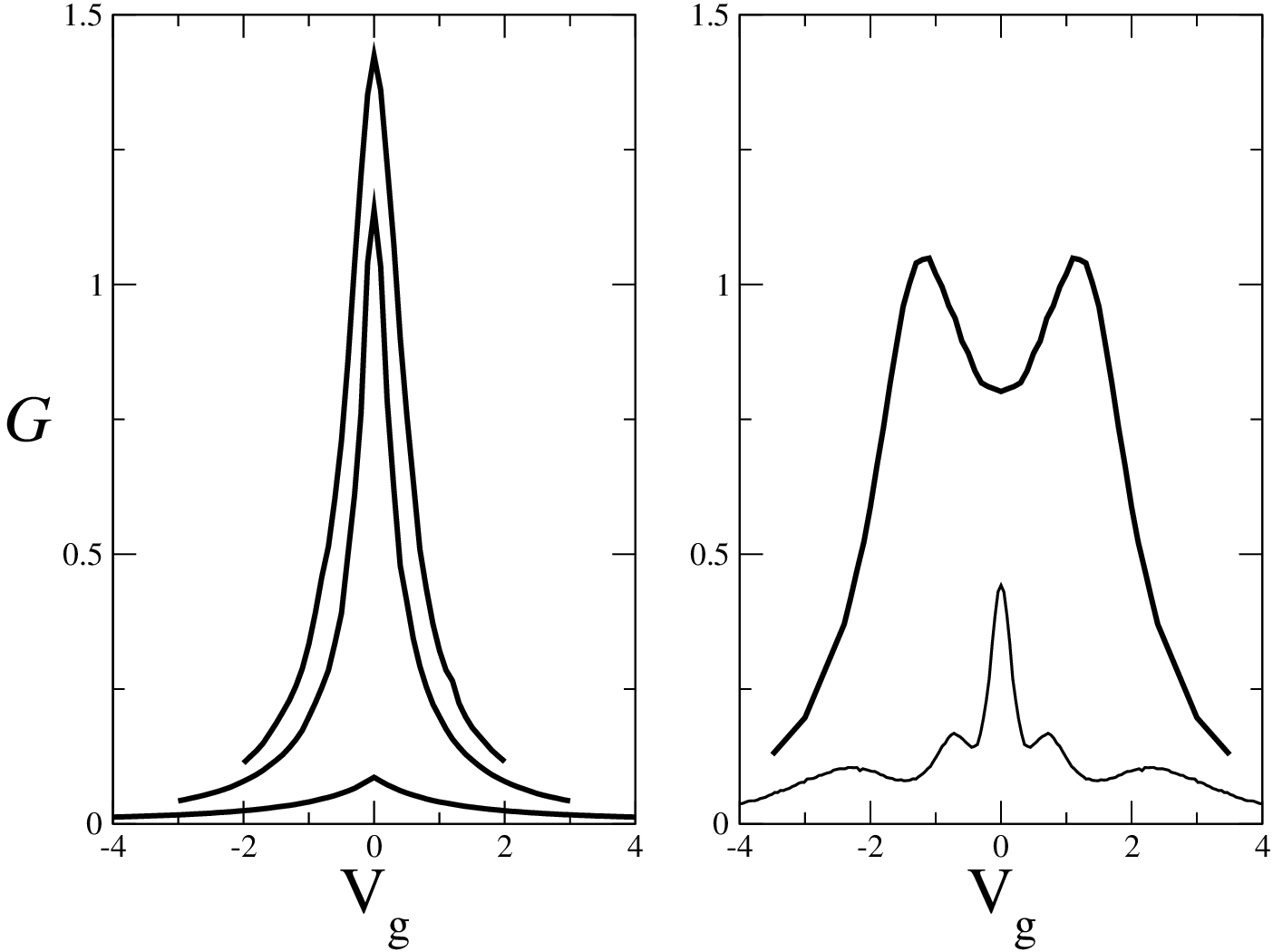}
\caption{Conductance (a. u.) at $T=0.0625$ and $\lambda=1$
as a function of the gate potential $V_g$.
Left: $U=0$ and $\Omega_0=0.333, 1, 2$  (bottom to top).
Right: $U=4$ and $\Omega_0=0.333$. For comparison, in thin 
line we show $\rho(\omega)$ for $V_g=0$.}
\label{fig4}
\end{figure}

We may now turn to the discussion of the consequences of the
above mentioned effects
on the conducting properties of this quantum impurity model
that should be relevant for nanoscopic devices such as 
a molecule attached to leads with both electronic and vibrational 
degrees of freedom.

One of the usual experimental setups to study transport phenomena
in nanostructures \cite{vibmol,konmol,kondot}, consists in 
applying a bias potential 
between the left and the right sides of the impurity (molecule), 
and then measuring the current through it. In addition
a gate voltage $V_g$ might be applied at the impurity site, and the
conductance is obtained as a function of $V_g$.

In the limit of very small bias between the left and right leads
the conductance can be calculated from \cite{miwi}
\begin{equation}
G=\frac{e}{2 \hbar}\sum_{\sigma}\int \frac{d \omega}{2\pi}
t_c^2 \rho_0(\omega)
\frac{\partial f(\omega)}{\partial \omega}
\rho(\omega),
\end{equation}
where $f(\omega)=1/(1+e^{\beta \omega})$.
Results at low $T$ are shown in Fig.\ref{fig4}. 
In this parameter regime, the behavior of $G$ is essentially
given by the value of $\rho(0)$.
For $U=0$ (left) a maximum in the
conductance is observed at $V_g=0$, and its value is strongly
enhanced as $U_{eff}$ is decreased. The maximal value of the conductance
is controlled by the spectral strength of the Kondo-like
resonance that is rapidly suppressed by $U_{eff}$. 
While the $V_g=0$ case is understood in
terms of the analogy to the usual Kondo case, we find that upon
application of a gate potential there is a very different
behavior with respect to the
case of conductance through a magnetic impurity. In the latter case, at
low $T$, the conductance displays peaks at $\pm U/2$ and remains sizable
in between \cite{kondot}.
 This is because the Kondo peak survives the departure
form the particle-hole symmetric situation, since $V_g$ does not
break the spin doublet at the impurity and the slow spin fluctuation
mechanism remains operational. In stark contrast, the present case shows
a rather narrow peak of the conductance centered at $V_g=0$ and this
is due to the fact that a finite $V_g$ immediately destroys the
empty - doubly occupied doublet at the impurity, and the
validity of the Kondo-like
analogy. At low temperatures, the probability of the participation
of those states differs by a factor $\sim exp(-V_g/T)$.
Switching on $U$ (right), we find a new surprise, namely that even for
$U_{dyn} < 0$ (ie, the phonon mediated interaction dominates)
the conductance remains high in a neighborhood of $V_g=0$
and develops a two peak
structure at $V_g \approx \pm U_{dyn}/2$. This behavior is consistent
with
a Kondo-like character that remains present even away from
particle hole symmetry. The systematic investigation of the conductance
in the whole parameter regime is left for future investigations.

In conclusion, we have introduced a new QMC algorithm for the numerical solution
of a general class of quantum impurities models with fermionic and bosonic
degrees of freedom. We considered the specific case of a SIAM with a Holstein
phonon that served both to benchmark the quality of our method and to demonstrate
new effects in the impurity conductance. This model and further generalizations
might be of great help for the investigation of conduction properties through
molecular nanodevices.
    
We thank Prof. Fulde for his hospitality and (LA) the support 
of the Alexander von Humboldt Stiftung. LA is currently at BIFI-Zaragoza
through RyC program of MCyT, Spain.  
We also thank useful discussions with D. Grempel
and P. Cornaglia and Y. Kakehashi. 
Support from CONICET and PICT 03-11609
of Argentina is also acknowledged.


\begin{thebibliography}{999}
\bibitem{hewsonbook}A.C. Hewson, {\sl The Kondo Problem to Heavy Fermions}, 
Cambridge Studies in Magnetism Vol. 2. Cambridge University Press, 
Cambridge, England (1993).

\bibitem{vibmol}H. Park et al, Nature {\bf 407}, 57 (2000).

\bibitem{konmol}J. Park et al, Nature {\bf 417}, 722 (2002);
W. Liang et al, Nature {\bf 417}, 725 (2002).

\bibitem{kondot} D. Goldhaber-Gordon et al, Nature {\bf 391}, 156 (1998).

\bibitem{hirfy}J. E. Hirsch and R. M. Fye, Phys. Rev. Lett. {\bf 56},
2521 (1986).

\bibitem{an} P. W. Anderson, Phys. Rev. {\bf 124}, 41 (1961).

\bibitem{noteqmc} In the actual implementation we use a similar
Dyson formula expression as in \cite{hirfy}. This simpler expression
is used here the sake of clarity.

\bibitem{hew} A. C. Hewson and D. Meyer, J. Phys. Cond. Mat. 
{\bf 14}, 427 (2002); D. Meyer, A. C. Hewson and R. Bulla,
Phys. Rev. Lett. {\bf 89}, 196401 (2002);
 P. S. Cornaglia, H. Ness, and D. R. Grempel 
Phys. Rev. Lett. {\bf 93}, 147201 (2004).

\bibitem{miwi} Y. Meir and N. S. Wingreen, Phys. Rev. Lett {\bf 68},
2512 (1992).
 
\bibitem{note}
While this may seem a drawback, in practice it turns out that
not visiting the high-$m$ bosonic fields usually does not degrade the
quality of the numerical results.

\end{thebibliography}
\end{document}